\documentclass{llncs}

\usepackage[british]{babel}
\usepackage[draft]{fixme}
\usepackage{booktabs}
\usepackage[list]{todo}
\usepackage{url}
\usepackage{xspace}
\usepackage{tabularx}
\usepackage{amsmath}
\usepackage{amssymb}
\usepackage{array}
\usepackage{color}
\usepackage{graphicx}
\usepackage{url}
\usepackage{times}
\usepackage{epsfig}
\usepackage{mathpartir}
\usepackage{listings}

\usepackage{subfigure}
\usepackage{tikz}
\usepackage{wrapfig}
\usepackage{desclist}

\makeatletter
\gdef\thelstlisting{\@arabic\c@lstlisting}
\makeatother

\usepackage{xspace}

\newcommand{\myfig}{Fig.~}
\newcommand{\mythm}{Thm.~}
\newcommand{\mylem}{Lem.~}

\newcommand{\mysec}{Sec.~}

\newcommand{\eg}{e.g.\xspace}
\newcommand{\ie}{i.e.\xspace}

\let\implies\Rightarrow

\let\iff\Leftrightarrow

  \newcommand{\dfnshort}[2]{{#1}\triangleq{#2}}
  
  \newcommand{\dfn}[2]{{#1}~\triangleq~{#2}}
  
  \newcommand{\type}[2]{{#1} : {#2}}

  \newcommand{\re}{r}

  \let\setset\setbf

  \newcommand{\evts}{\setset{E}}

  \newcommand{\lo}{\ell}
  \newcommand{\loc}{\operatorname{addr}}

  \newcommand{\VX}{\operatorname{valid}}

\newcommand{\stacklabel}[1]
{\stackrel{\smash{\scriptstyle\textnormal{#1}}}}

  \newcommand{\po}{\textsf{po}}

  \newcommand{\dep}{\textsf{dp}}

  \newcommand{\rf}{\textsf{rf}}

  \newcommand{\rfe}{\textsf{rfe}}

  \newcommand{\rfi}{\textsf{rfi}}

  \newcommand{\ws}{\textsf{ws}}

  \newcommand{\fr}{\textsf{fr}}

  \newcommand{\fenced}{\textsf{fence}}
  
  \newcommand{\poi}[1]{\textsf{po\ensuremath{_{#1}}}}
    
  \newcommand{\pio}{\operatorname{\textsf{po-loc}}}

  \newcommand{\acyclic}{\operatorname{acyclic}}
  
  \newcommand{\uniproc}{\operatorname{uniproc}}
  
  \newcommand{\thin}{\operatorname{thin}}
  \newcommand{\valid}{\operatorname{valid}}
  
  \newcommand{\consensus}{\operatorname{consensus}}

    \newcommand{\cdelays}{\operatorname{crits}}
    \newcommand{\pdelays}{\operatorname{delays}}

\newcommand{\import}{\operatorname{import}}

\let\prog\textsf
\let\as\texttt
\let\ltest\textbf

\newcommand{\proc}[1]{\ensuremath{P_{#1}}}
\newlength{\fmtlength}

\newcommand{\lb}[4]{\((#1)\)\,#2#3#4}

\newcommand{\Cands}{relations}

\newcommand{\llb}[3]{#1#2#3}
\newcommand{\safes}{\operatorname{safe}}

\newcommand{\DR}{\operatorname{r}}
\newcommand{\DW}{\operatorname{w}}
\newcommand{\Del}{\operatorname{d}}
\newcommand{\Flush}{\operatorname{f}}
\newcommand{\Mem}{\operatorname{m}}

\newcommand{\wpath}{\operatorname{path}}

\newcommand{\mns}{\operatorname{mns}}
\newcommand{\dpo}{{\bf (dpo)}}
\newcommand{\drfs}{{\bf (drfs)}}
\newcommand{\drft}{{\bf (drft)}}

\newcommand{\lw}{\operatorname{\textsf{l}}}
\newcommand{\mem}{\operatorname{\textsf{m}}} 
\newcommand{\buff}{\operatorname{\textsf{b}}} 
\newcommand{\queue}{\operatorname{\textsf{q}}} 
\newcommand{\state}{\operatorname{\textsf{s}}} 
\newcommand{\ue}{\textsf{de}}
\newcommand{\se}{\textsf{se}}

\newcolumntype{Y}{@{}r@{\,}X}
\newcommand{\instab}[2]{\ \(#2\)  & \as{#1}}

\newcommand{\pset}[2]{\(\as{#1} \leftarrow \as{#2}\)}
\newcommand{\pstore}[2]{\pset{#2}{#1}}
\newcommand{\pload}[2]{\pset{#1}{#2}}

\newcommand{\plwarx}[3]{\as{lwarx #1,#2,#3}}
\newcommand{\pstwcx}[3]{\as{stwcx. #1,#2,#3}}
\newcommand{\pstw}[3]{\as{stw #1,#2,#3}}
\newcommand{\pbne}[1]{\as{bne #1}}
\newcommand{\pcmp}[2]{\as{cmpw #1,#2}}
\newcommand{\haut}{\rule{0ex}{2ex}}
\newcommand{\bas}{\rule[-1ex]{0.5ex}{0ex}}

\newcommand{\Rmw}[1][40ex]
{\begin{tabular}{rl}
\haut \instab{\as{loop:}}{} \\
\instab{\plwarx{r1}{0}{r5}}{(a_1)} \\
\instab{[\dots]}{}\\
\instab{\pstwcx{r2}{0}{r5}}{(a_2)} \\
\bas\instab{\pbne{loop}}{(b)} \\ 
\end{tabular}}
\newcommand{\Atom}[1][40ex]
{\begin{tabular}{rl}
\haut \instab{\plwarx{r1}{0}{r5}}{(a_1)} \\
\instab{[\dots]}{}\\
\bas\instab{\pstwcx{r2}{0}{r5}}{(a_2)} \\
\end{tabular}}
\newcommand{\Lo}[1][40ex]
{\begin{tabular}{rl}
\haut\instab{\as{loop:}}{} \\
\instab{\plwarx{r6}{0}{r3}}{(a_1)} \\
\instab{\pcmp{r4}{r6}}{(b)} \\
\instab{\pbne{loop}}{(c)} \\
\instab{\pstwcx{r5}{0}{r3}}{(a_2)} \\
\instab{\pbne{loop}}{(d)} \\
\instab{\as{isync}}{(e)} \\
\bas\instab{[\dots]}{}\\ 
\end{tabular}}
\newcommand{\ULo}[1][40ex]
{\begin{tabular}{rl}
\instab{[\dots]}{} \\
\haut\instab{\as{lwsync}}{(f)} \\ 
\bas\instab{\pstw{r4}{0}{r3}}{(g)} \\
\end{tabular}}
\newcommand{\Iriw}[1][60ex]
{\begin{tabularx}{#1}{Y|Y|Y|Y}
\multicolumn{8}{c}{\ltest{iriw}}\\ \hline
\multicolumn{2}{c|}{\haut\proc{0}} &
\multicolumn{2}{c|}{\proc{1}} &
\multicolumn{2}{c|}{\proc{2}} &
\multicolumn{2}{c}{\proc{3}} \\ \hline
\instab{\pload{r1}{x}}{(a)}\haut &
\instab{\pload{r3}{y}}{(c)} &
\instab{\pstore{1}{x}}{(e)} &
\instab{\pstore{1}{y}}{(f)}\\
\bas\instab{\pload{r2}{y}}{(b)} &
\instab{\pload{r4}{x}}{(d)} &
\instab{}{} &
\instab{}{}
\\ \hline
\multicolumn{8}{l}{\haut{}Allowed? \as{r1=1}; \as{r2=0}; \as{r3=1};
\as{r4=0};}
\end{tabularx}}
\newcommand{\AOE}[1][40ex]
{\begin{tabularx}{#1}{Y|Y}
\multicolumn{4}{c}{\ltest{sb}} \\ \hline
\multicolumn{2}{c|}{\haut\proc{0}} &
\multicolumn{2}{c}{\haut\proc{1}} \\ \hline
\haut\instab{\pstore{1}{x}}{(a)} & \instab{\pstore{1}{y}}{(c)} \\
\bas\instab{\pload{r1}{y}}{(b)} & \instab{\pload{r2}{x}}{(d)} \\ \hline 
\multicolumn{4}{l}{Allowed? \as{r1=0}; \as{r2=0} } \\
\end{tabularx}}

\makeatletter
\let\UrlSpecialsOld\UrlSpecials
\def\UrlSpecials{\UrlSpecialsOld\do\/{\Url@slash}\do\_{\Url@underscore}}%
\def\Url@slash{\@ifnextchar/{\kern-.11em\mathchar47\kern-.2em}%
    {\kern-.0em\mathchar47\kern-.08em\penalty\UrlBigBreakPenalty}}
\def\Url@underscore{\nfss@text{\leavevmode \kern.06em\vbox{\hrule\@width.3em}}}
\makeatother


\title{Software Verification for Weak Memory via Program
Transformation\thanks{Supported by EPSRC project~EP/G026254/1 and
the Semiconductor Research Coropration (SRC) under task~2269.002.}}

\author{Jade Alglave \and Daniel Kroening \and Vincent Nimal \and Michael Tautschnig}

\institute{Department of Computer Science, University of Oxford, UK}

\begin{document}
\maketitle


\begin{abstract}
Despite multiprocessors implementing weak memory models, verification
methods often assume \emph{Sequential Consistency} (SC), thus may miss bugs due
to weak memory.  We propose a sound transformation of the program to
verify, enabling SC tools to perform verification w.r.t.~weak memory.
We present experiments for a broad variety of models (from x86/TSO to
Power/ARM) and a vast range of verification tools, quantify the additional
cost of the transformation and highlight the cases when we can
drastically reduce it. Our benchmarks include work-queue management
code from PostgreSQL.
%
\end{abstract}


\section{Introduction}

Current multi-core architectures such as Intel's x86, IBM's Power or ARM,
implement \emph{weak memory models} for performance reasons, allowing
optimisations such as \emph{instruction reordering}, \emph{store buffering}
or \emph{write atomicity relaxation}~\cite{ag96}.  These models make
concurrent programming and debugging extremely challenging, because the
execution of a concurrent program might not be an interleaving of its
instructions, as would be the case on a Sequentially Consistent (SC)
architecture~\cite{lam79}.
%
As an
instance, the lock-free signalling code in the open-source database
PostgreSQL failed on regression tests on a PowerPC cluster, due to the
memory model.  We study this bug in detail in
\mysec\ref{experiments}.

This observation highlights the crucial need for weak memory aware
verification.  Yet, most existing work assume SC~\cite{rin01}, hence might miss
bugs specific to weak memory.  Recent work addresses the design or the
adaptation of existing methods and tools to weak
memory~\cite{pd95,ygl04,hr06,bam07,owe10,abp11,aac12}, but often focuses on one
specific model or cannot handle the write atomicity relaxation of
Power/ARM: generality remains a challenge.

Since we want to avoid writing one tool per architecture of interest, we
propose a unified method.  Given a program analyser handling SC concurrency
for C programs, we \emph{transform its input} to simulate the possible
non-SC behaviours of the program whilst executing the program on SC.
Essentially, we augment our programs with arrays to simulate (on SC) the
buffering and caching scenarios due to weak memory.

The verification problem for weak memory models is known to be hard (\eg
non-primitive recursive for TSO), if not undecidable (\eg for RMO-like
models)~\cite{abb10}. In practice, this means that we cannot design a
\emph{complete} verification method.  Yet, we can achieve \emph{soundness}, by
implementing our tools in tandem with the design of a proof, and by stressing
our tools with test cases reflecting subtle points of the proof.

We also aim for an effective and unified verification setup, where one can
easily plug a tool of choice. This paper meets these objectives by making
three new contributions:
\begin{enumerate}

\item \mysec\ref{instrumentation} details our \emph{transformation}
for concurrent programs on weak memory. This requires defining a
generic abstract machine that we prove (in the Coq proof assistant) equivalent
to the framework of~\cite{ams10} (recalled in \mysec\ref{context}).
\mysec\ref{instrumentation} shows a drastic
optimisation of the transformation, and we prove that this is sound.

\item \mysec\ref{implementation} describes our implementation, where the
generality of our approach reveals itself the most: we support a broad variety
of models (x86/TSO, PSO, RMO and Power) and program analysers
(Blender~\cite{kvy11}, CheckFence~\cite{bam07},
ESBMC~\cite{DBLP:conf/icse/CordeiroF11}, MMChecker~\cite{hr06},
Poirot~\cite{poirot}, SatAbs~\cite{dkkw2011-cav}, Threader~\cite{gpr11}, and
our new CImpact tool, an extension of Impact~\cite{DBLP:conf/cav/McMillan06} to
SC concurrency).

\item \mysec\ref{experiments} details our experiments using this setup. 
i) We systematically validate our implementation w.r.t.~our
theoretical study with $555$ \emph{litmus tests}, generated by
the \prog{diy} tool~\cite{ams10} to exercise weak memory artefacts
in isolation.
ii) We verify several TSO examples from the literature
\cite{dij65,pet81,lam87,szy88,bm08:cav}.  
iii) We verify a new example, which is an excerpt of the
relational database software PostgreSQL and has a bug specific to Power. This
bug raised notable interest at IBM, and we are already trying our tools on
their software.
\end{enumerate}

We provide the source and documentation of our tools, our benchmarks, Coq
proofs and experimental reports
online: \url{www.cs.ox.ac.uk/people/vincent.nimal/instrument/}

\paragraph{Related Work}

We focus here on the \emph{verification} problem, \ie forbidding
the behaviours that are buggy, not all the non-SC ones.  This problem
is non-primitive recursive for TSO~\cite{abb10}.  It is undecidable if the
reads are smart (\ie they can guess the value that they will read eventually),
\eg for RMO-like models~\cite{abb10}. Forbidding \emph{causal loops} restores
decidability; 
relaxing write atomicity makes the problem undecidable again
\cite{abb12}.

Previous work therefore compromise by choosing various
bounds over the objects of the model~\cite{abp11,kvy10}, over-approximating the
possible behaviours~\cite{kvy11,jys12}, or relinquishing
termination~\cite{lw11}.  For TSO,~\cite{aac12} presents a sound and complete
solution. 

By contrast, we disregard in the present paper any completeness issue. We are
not primarily concerned with efficiency either, although we do provide a
drastic optimisation of our transformation.  
focus in this work on the soundness, generality, and implementability of our
method, to bridge the gap between theory and practice. We emphasise the fact
that our method allows to lift any SC method or tool to a large spectrum of
weak memory models, ranging from x86 to Power.



\section{Context: Axiomatic Model\label{context}}

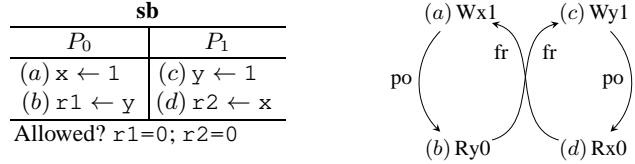
\begin{figure}[!t]
\begin{tabular}{m{.5\linewidth}m{.5\linewidth}}
\centerline{\AOE[.6\linewidth]}
&
  \scalebox{0.9}{
  \begin{tikzpicture}[>=stealth,thin,inner sep=0pt,text centered,shape=rectangle]
  \useasboundingbox (-1cm,0.5cm) rectangle (2cm,-2.6cm);
  
    \begin{scope}[minimum height=0.5cm,minimum width=0.5cm,text width=1.0cm]
      \node (a)  at (0, 0)  {\lb aWx{1}};
      \node (b)  at (0, -2)  {\lb bRy{0}};
      \node (c)  at (2, 0)  {\lb cWy{1}};
      \node (d)  at (2, -2)  {\lb dRx{0}};
    \end{scope}
		
    \path[->] (a) edge [out=225,in=135] node [left=0.1cm] {po} (b);
    \path[->] (b) edge [out=15,in=195] node [pos=0.8, below right=0.07cm] {fr} (c);
    \path[->] (c) edge [out=-45,in=45] node [left=0.1cm] {po} (d);
    \path[->] (d) edge [out=165,in=-15] node [pos=0.8, below left=0.07cm] {fr} (a);
  \end{tikzpicture}}
\end{tabular}
\vspace*{-8mm}
\caption{\label{fig:sb} Store Buffering (\ltest{sb})}
\vspace*{-4mm}
\end{figure}

We use the framework of~\cite{ams10}, which provably embraces several
\emph{architectures}: SC~\cite{lam79}, Sun TSO (\ie the x86
model~\cite{oss09}), PSO and RMO
, Alpha
, and a
fragment of Power
. We present this framework 
via \emph{litmus tests}, as shown in \myfig\ref{fig:sb}. 

The keyword \emph{allowed} asks if a given architecture allows the outcome
``{\tt r1=0;r2=0}''.  This relates to the execution graphs of this program,
composed of relations over \emph{read and write memory events}.  A store
instruction (\eg \pstore{1}{x} on $P_0$) corresponds to a write event
(\llb{$(a)$}{W}{x}{$1$}), and a load (\eg \pload{r1}{y} on $P_0$) to a read
(\llb{$(b)$}{R}{y}{0}).  The validity of an execution boils down to the absence
of certain cycles in the execution graph.  Indeed, an architecture allows an
execution when it represents a \emph{consensus} amongst the processors.  A
cycle in an execution graph is a potential violation of this consensus.

If an execution graph has a cycle, we check if the architecture \emph{relaxes}
some relations in this cycle. The consensus can ignore a relaxed relation,
hence become acyclic, \ie the architecture allows the final state. In
\myfig\ref{fig:sb}, on SC where no relation is relaxed, the cycle forbids the
execution. x86 relaxes the program order ($\po$ in \myfig\ref{fig:sb})
between writes and reads, thus a forbidding cycle no longer
exists since $(a,b)$ and $(c,d)$ are relaxed.

\paragraph{Executions}
\begin{figure}[!t]
\begin{tabular}{m{.6\linewidth}@{}m{.4\linewidth}}
\centerline{\Iriw[\linewidth]}
&
  \scalebox{0.7}{
  \begin{tikzpicture}[>=stealth,thin,inner sep=0pt,text centered,shape=rectangle]
  \useasboundingbox (-1.5cm,1cm) rectangle (3.5cm,-3.2cm);
  
    \begin{scope}[minimum height=0.5cm,minimum width=0.5cm,text width=1.0cm]
      \node (a)  at (0, 0)  {\lb aRx{1}};
      \node (b)  at (0, -2)  {\lb bRy{0}};
      \node (c)  at (1.5, 0)  {\lb cRy{1}};
      \node (d)  at (1.5, -2)  {\lb dRx{0}};
      \node (e)  at (3.0, 0)  {\lb eWx{1}};
      \node (f)  at (4.5, 0)  {\lb fWy{1}};
    \end{scope}
		
    \path[->] (a) edge [out=225,in=135] node [left=0.1cm] {po} (b);
    \path[->] (c) edge [out=225,in=135] node [left=0.1cm] {po} (d);
    \path[->] (e) edge [out=150,in=30] node [above=0.1cm] {rf} (a);
    \path[->] (b) edge [out=-30,in=225,looseness=1.0] node [below right=0.07cm] {fr} (f);
    \path[->] (f) edge [out=150,in=30] node [above=0.1cm] {rf} (c);
    \path[->] (d) edge [out=30,in=-120] node [above left=0.07cm] {fr} (e);
  \end{tikzpicture}}
\end{tabular}
\vspace*{-8mm}
\caption{\label{fig:iriw} Independent Reads of Independent Writes (\ltest{iriw})}
\vspace*{-4mm}
\end{figure}
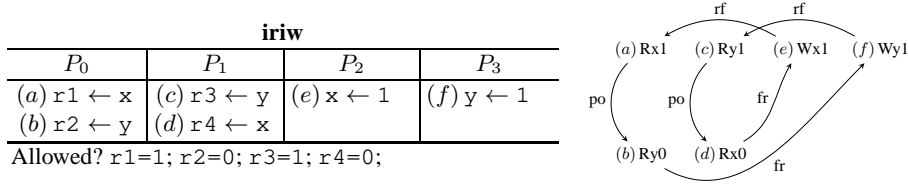
Formally, an \emph{event} is a read or a write memory access, composed of a
unique identifier, a direction R for read or W for write, a memory address,
and a value. We represent each instruction by the events it issues.
In~\myfig\ref{fig:iriw}, we associate the store \pstore{1}{$x$} on processor
$P_2$ to the event~\llb{$(e)$}{W}{$x$}{$1$}.   

We associate the program with an \emph{event structure} $\dfnshort{E}{(\evts,\po)}$,
composed of its events $\evts$ and the \emph{program order} $\po$, a
per-processor total order. We write $\dep$ for the relation (included in $\po$,
the source being a read) modelling \emph{dependencies} between instructions,
$\eg$ an \emph{address dependency} occurs when computing the address of a load
or store from the value of a preceding load. 
 
Then, we represent the \emph{communication} between processors leading to the
final state via an \emph{execution witness} $\dfnshort{X}{(\ws,\rf)}$, which
consists of two relations over the events.  First, the \emph{write
serialisation} $\ws$ is a per-address total order on writes which models the
\emph{memory coherence} widely assumed by modern
architectures
. It links a write $w$ to any write $w'$ to the same address that hits the
memory after $w$.  Second, the \emph{read-from} relation $\rf$ links a
write~$w$ to a read~$\re$ such that $\re$ reads the value written by~$w$. 
 
We include the writes in the consensus via the write serialisation.
Unfortunately, the read-from map does not give us enough information to embed
the reads as well. To that aim, we derive the \emph{from-read} relation $\fr$
from $\ws$ and $\rf$.  A read $r$ is in $\fr$ with a write $w$ when the write
$w'$ from which $r$ reads hit the memory before $w$ did.  Formally, we have:
$\dfn{(r, w) \in \fr}{\exists w', (w',r) \in \rf \wedge (w',w) \in \ws}$.
 
In \myfig\ref{fig:iriw}, the specified outcome corresponds to the execution on
the right if each memory location and register initially holds $0$. If {\tt
r1=1} in the end, the read $(a)$ read its value from the write $(e)$ on $P_2$,
hence $(e,a) \in \rf$. If {\tt r2=0} in the end, the read $(b)$ read its value
from the initial state, thus before the write $(f)$ on $P_3$, hence $(b,f) \in
\fr$.  Similarly, we have $(f,c) \in \rf$ from {\tt r3=1}, and $(d,e) \in \fr$
from {\tt r4=0}. 

\paragraph{Relaxed or safe}

A processor can commit a write $w$ first to a store buffer, then to a cache,
and finally to memory. When a write hits the memory, all the processors agree
on its value. But when the write $w$ transits in store buffers and caches, a
processor can read its value through a read $r$ before the value is actually
available to all processors from the memory. In this case, the read-from
relation between the write $w$ and the read $r$ does not contribute to the
consensus, since the reading occurs in advance.  

We model this by some subrelation of the read-from $\rf$ being \emph{relaxed},
\ie not included in the consensus. When a processor can read from its own
store buffer~\cite{ag96} (the typical TSO/x86 scenario), we relax the internal
read-from $\rfi$. When two processors $P_0$ and $P_1$ can communicate privately
via a cache (a case of \emph{write atomicity} relaxation~\cite{ag96}),
we relax the external read-from $\rfe$, and call the corresponding write
\emph{non-atomic}. This is the main particularity of Power or ARM, and cannot
happen on TSO/x86.

Some program-order pairs may be relaxed (\eg write-read pairs on x86, and
all but \dep~ones on Power), \ie only a subset of $\po$ is
guaranteed to occur in this order. 
%
Architectures provide special \emph{fence} (or \emph{barrier}) instructions,
to prevent weak behaviours.  
Following~\cite{ams10}, the relation
${\fenced} \subseteq {\po}$ induced by a fence is \emph{non-cumulative} when it
orders certain pairs of events surrounding the fence, \ie $\fenced$ is safe.
The relation $\fenced$ is \emph{cumulative} when it makes writes atomic, \eg
by flushing caches. The relation $\fenced$ is \emph{A-cumulative}
(resp.~\emph{B-cumulative}) if $\rfe;\fenced$ (resp.~$\fenced;\rfe$) is
safe. 
When stores are atomic (\ie $\rfe$ is safe), \eg on 
TSO, we do not need cumulativity.

\paragraph{Architectures}
An \emph{architecture} $A$ determines the set $\safes_A$ of the \Cands~safe on
$A$, \ie the relations embedded in the consensus.  Following \cite{ams10}, we
consider the write serialisation $\ws$ and the from-read relation $\fr$ to be
always \emph{safe}, \ie not relaxed. 
SC relaxes nothing, \ie $\rf$ and $\po$ are safe.  TSO authorises the
reordering of write-read pairs and store buffering
(\ie $\poi{\textsf{WR}}$ and $\rfi$ are relaxed) but nothing else. 

Finally, an execution $(E,X)$ is \emph{valid} on $A$ when the three following
conditions hold.
1.~SC holds per address, \ie the communication 
and the program order for accesses with same address $\pio$ are compatible:
$\uniproc(E,X) \triangleq {\acyclic(\ws \cup \rf \cup \fr \cup \pio)}$. 
2.~Values do not come out of thin air, \ie there is no causal loop:
$\thin(E,X) \triangleq {\acyclic(\rf \cup \dep)}$.
3.~There is a consensus, \ie
the safe \Cands~do not form a cycle: $\consensus(E,X) \triangleq
{\acyclic({({\ws \cup \rf \cup \fr} \cup {\po})} \cap {\safes_A})}$.  
Formally:
\[\VX_{A}(E,X) \triangleq \uniproc(E,X) \wedge \thin(E,X) \wedge \consensus(E,X) \]

\vspace{-1ex}

\section{Simulating Weak Behaviours on SC \label{instrumentation}}

We want to transform a program $P$ into a program $P'$ so that executing $P'$
on SC gives us the behaviours that $P$ exhibits on a weak architecture.  To do
so, we define an \emph{abstract machine}, composed of a store
buffer per address and a load queue.  We avoid defining one machine per
architecture as follows.  We first define a \emph{core machine}, implementing
the $\uniproc$ and $\thin$ checks common to all models.  Then, we implement an
architecture $A$ by adding, on top of the core, a \emph{protocol} ordering the
entering and exiting the buffers and queue.  This protocol enforces the
consensus order defined by~$A$.

\subsection{The Abstract Machine}
\label{sec:full}

A \emph{state} $s$ is either $\bot$ or a tuple $(\mem,\buff,\queue,\lw)$, which
contains (writing $\textbf{addr}, \textbf{evt}, \textbf{proc}, \textbf{rln}$ for the
types of memory addresses, events, processors (or thread ids) and relations):
\begin{description}
\item[\small{the \emph{memory} $\type{\mem}{\text{addr} \rightarrow \text{evt}}$}] 
maps a memory address $\lo$ to a write to $\lo$;
\item[\small{a \emph{buffer} $\type{\buff}{\text{addr} \rightarrow \text{rln evt}}$}]
a total order over writes to the same address;
\item[\small{a \emph{queue} $\type{\queue}{\text{rln (evt $\times$ evt)}}$}]  
over reads, tracking instruction dependencies;
\item[\small{the \emph{log} $\type{\lw}{\text{proc} \times \text{addr}
\rightarrow \text{evt}}$}] provides the
last write to address $\lo$ seen by thread $p$.
\end{description}

Note that our buffer is not a per-thread object, but solely a per-location
object, as opposed to most existing formalisations. This allows us to model not
only store buffering (which per-thread objects would allow), but also caching
scenarios as exhibited by \ltest{iriw+dps} (\ie the \ltest{iriw} test of
\myfig\ref{fig:iriw} with dependencies between the reads on $P_0$ and $P_1$ to
prevent their reordering), \ie fully non-atomic stores. 

\paragraph{Core machine}

We modify a state \emph{via} \emph{labels}.  Given an execution $(E,X)$, we
define our labels from the events of $E$.  First, we \emph{augment} our events:
a write $w$ becomes $\DW(w)$ and $r$ becomes $\DR(w,r)$, where $w$ is the write
from which $r$ reads (\ie $(w,r) \in \rf$).  Then we tag an augmented event $e$
to build the labels $\Del(e)$ and $\Flush(e)$. In effect, we \emph{split} an
event $e$ into its \emph{delayed} part $\Del(e)$ (the part entering the
buffer or the queue), and its \emph{flushed} part $\Flush(e)$ (the part
exiting the buffer or the queue).

A label modifies a state $\state=(\mem,\buff,\queue,\lw)$ as follows.  We
describe the transitions in prose, and omit their formal definitions for
brevity (our Coq proofs are available online):
\begin{description}
\item[Write to buffer] a write $\Del(\DW(w))$ to location $\lo$ can always enter
the buffer $\buff$, taking its place after all the writes to $\lo$ that are
already in $\buff$;
\item[Write from buffer to memory] a write $\Flush(\DW(w))$ to location $\lo$ and
from thread $p$ exits the buffer $\buff$ and updates the memory to $\lo$ if there is
no write to $\lo$ pending in $\buff$, and if there is no pending read from
$\lo$ enqueued by $p$ before $w$ entered $\buff$;
\item[Enqueue read] a read $\Del(\DR(w,r))$ can always enter the queue $\queue$,
taking its place after all the reads in $\queue$ on which $r$ depends (\ie
$\{r' \mid (r',r) \in \dep \vee (r',w) \in \dep)$);
\item[Read from queue] a read $\Flush(\DR(w,r))$ by thread $p$ from $\lo$ exits
the queue and updates the log relative to $p$ and $\lo$ if there is no pending
read on which $r$ depends; if no write to $\lo$ and from $p$ entered the buffer
$\buff$ after (resp.~before) $r$ yet took its place in $\buff$ before
(resp.~after) $w$; and if $w$ took its place in $\buff$ after the last write to $\lo$
seen by $p$.
\end{description}

The rules for writes enforce the existence of a write serialisation, since we
order the writes to a given location in the buffer, then flush them in the same
order. Reads are smart (to the extent that they respect $\uniproc$ or
$\thin$): we flush a read $\Flush(\DR(w,r))$ if the write $w$ lies in
the memory or in the buffer, \ie~in any level of the memory hierarchy.

Finally, note that the core machine implements the $\uniproc$ and $\thin$
checks that all architectures satisfy, thanks to the premises to exiting the
buffer or the queue.

\paragraph{Consensus protocol}
Now, to implement a particular architecture $A$, the machine also needs to
implement the consensus defined by $A$.  We do so by defining a protocol that
constrains the order in which reads and writes exit the queue and the buffer,
as follows.

We first gather the \emph{delay pairs} of $A$ (adapting the terminology
of~\cite{ss88,am11}).  The delays make a program behave differently on $A$ than
on SC, as follows~\cite{ams10,am11}.  First, when the program has a relaxed
program order pair, \eg $(a,b)$ in \myfig\ref{fig:sb} on TSO.  Second, when the
program reads from a non-atomic write, \eg $(e,a)$ in \myfig\ref{fig:iriw} on
Power. Formally:
\begin{align*} 
\pdelays_{A}(E,X) & \triangleq \{ (e,e') \in (\po \cup \rfe) \wedge (e,e') \not\in \safes_A \}
 \end{align*}

Note that there is no delay on SC, and that the $\rfe$ case only concerns architectures
relaxing write atomicity, \eg Power/ARM, but not x86. In practice, the
delays are:
%
the write-read pairs on x86/TSO, \eg in \myfig\ref{fig:sb} $(a,b)$ and $(c,d)$;
the write-read and write-write pairs on PSO;
all \po~pairs (except \dep~ones) on RMO and Power;
all \rfe~pairs, on Power, \eg in \myfig\ref{fig:iriw} $(e,a)$ and $(f,c)$. 

In the following, $e$ stands for both the event $e$ and its augmented event.
To implement the consensus order defined by a given architecture $A$, we
\emph{augment our core machine with an $A$-protocol}.  Formally, we feed it a
path of labels $\wpath(E,X,\mathbb{D})$ defined as follows (where $\mathbb{D}$
is a set of pairs to be delayed):
\begin{description}
\item[\label{Ppo}Enter in \po] we ensure that our machine has an SC
semantics, by forcing two events $e_1$ and $e_2$ in program order to enter the
buffer and the queue in this order;
 \ie $(e_1,e_2) \in \po \implies (\Del(e_1),\Del(e_2)) \in \wpath(E,X,\mathbb{D})$;
\item[\label{Pghb}Safe exit (\se)] if $(e_1,e_2)$ does not form a
delay, we force them to exit the buffer and the queue in the same order
\ie $(e_1,e_2) \not\in \mathbb{D} \implies  (\Flush(e_1),\Flush(e_2)) \in \wpath(E,X,\mathbb{D})$;
\item[\label{Prf}Delayed exit (\ue)] if $(e_1,e_2)$ forms a delay, we force them to exit the buffer and the queue in the converse order
\ie $(e_1,e_2) \in \mathbb{D} \implies  (\Flush(e_2),\Flush(e_1)) \in \wpath(E,X,\mathbb{D})$.
\end{description}

Observe that if there are no delay pairs (\ie $\mathbb{D}=\emptyset$), this definition
describes all possible interleavings, \ie our machine implements SC:
\begin{lemma}
\label{lem:sc_equiv}$\valid_{\text{SC}}(E,X) \iff \mns(E,\wpath(E,X,\emptyset))$
\end{lemma}

\paragraph{Examples} We illustrate how the
machine implements TSO or Power by revisiting the \ltest{sb} test of
\myfig\ref{fig:sb} for TSO and the \ltest{iriw} test of \myfig\ref{fig:iriw}
for Power.

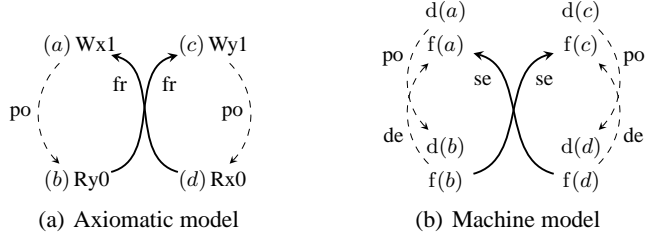
\begin{figure}[!t]
\centering
  \subfigure[Axiomatic model]{
  \scalebox{0.9}{
  \begin{tikzpicture}[>=stealth,thin,inner sep=0pt,text centered,shape=rectangle]
  \useasboundingbox (-1cm,0.5cm) rectangle (3cm,-2.25cm);
  
    \begin{scope}[minimum height=0.5cm,minimum width=0.5cm,text width=1.0cm]
      \node (a)  at (0, 0)  {\lb aWx{1}};
      \node (b)  at (0, -2)  {\lb bRy{0}};
      \node (c)  at (2, 0)  {\lb cWy{1}};
      \node (d)  at (2, -2)  {\lb dRx{0}};
    \end{scope}
		
    \path[->] (a) edge [out=225,in=135,dashed] node [left=0.1cm] {po} (b);
    \path[->] (b) edge [out=15,in=195,thick] node [pos=0.8, below right=0.07cm] {fr} (c);
    \path[->] (c) edge [out=-45,in=45,dashed] node [left=0.1cm] {po} (d);
    \path[->] (d) edge [out=165,in=-15,thick] node [pos=0.8, below left=0.07cm] {fr} (a);
  \end{tikzpicture}}}
  \hspace*{1cm}
  \subfigure[Machine model]{
  \scalebox{0.9}{
  \begin{tikzpicture}[>=stealth,thin,inner sep=0pt,text centered,shape=rectangle]
  \useasboundingbox (-1cm,0.5cm) rectangle (3cm,-2.75cm);
  
    \begin{scope}[minimum height=0.5cm,minimum width=0.5cm,text width=0.8cm]
      \node (da)  at (0, 0)  {$\Del(a)$};
      \node (fa)  at (0, -0.5)  {$\Flush(a)$};
      \node (db)  at (0, -2)  {$\Del(b)$};
      \node (fb)  at (0, -2.5)  {$\Flush(b)$};
      \node (dc)  at (2, 0)  {$\Del(c)$};
      \node (fc)  at (2, -0.5)  {$\Flush(c)$};
      \node (dd)  at (2, -2)  {$\Del(d)$};
      \node (fd)  at (2, -2.5)  {$\Flush(d)$};
    \end{scope}
		
    \path[->] (da) edge [out=225,in=135,dashed] node [pos=0.3,left=0.1cm] {po} (db);
    \path[->] (fb) edge [out=135,in=225,dashed] node [pos=0.3,left=0.1cm] {de} (fa);
    \path[->] (fb) edge [out=15,in=195,thick] node [pos=0.8, below right=0.07cm] {se} (fc);
    \path[->] (dc) edge [out=-45,in=45,dashed] node [pos=0.3,right=0.1cm] {po} (dd);
    \path[->] (fd) edge [out=45,in=-45,dashed] node [pos=0.3,right=0.1cm] {de} (fc);
    \path[->] (fd) edge [out=165,in=-15,thick] node [pos=0.8, below left=0.07cm] {se} (fa);
  \end{tikzpicture}}}
\caption{\label{fig:sb-df} Revisiting \ltest{sb} with our core machine augmented with a TSO protocol}
\vspace*{-1em}
\end{figure}

In \myfig\ref{fig:sb}, the pairs $(a,b)$ on $P_0$ and $(c,d)$ on $P_1$ are
delays on TSO. Our machine simulates the weak behaviour exhibited on
TSO, following the scenario in \myfig\ref{fig:sb-df}. The machine
buffers or enqueues all events w.r.t.~program order.  Since $(a,b)$ and $(c,d)$
are delays on TSO, the machine augmented with a TSO protocol flushes
the reads $b$ and $d$ before the writes $a$ and $c$, ensuring that the
registers {\tt r1} and {\tt r2} hold $0$ in the end.

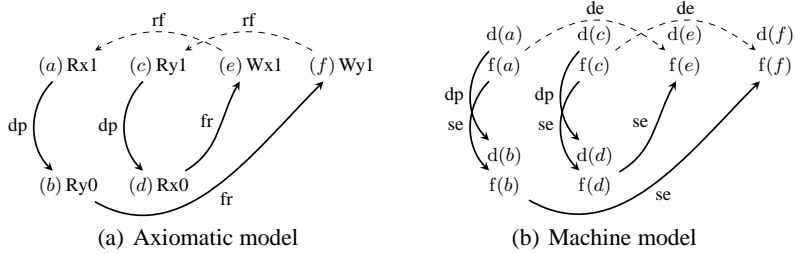
\begin{figure}[!t]
\centering
  \subfigure[Axiomatic model]{
  \scalebox{0.8}{
  \begin{tikzpicture}[>=stealth,thin,inner sep=0pt,text centered,shape=rectangle]
  \useasboundingbox (0cm,1cm) rectangle (4.5cm,-2.5cm);
  
    \begin{scope}[minimum height=0.5cm,minimum width=0.5cm,text width=1.0cm]
      \node (a)  at (0, 0)  {\lb aRx{1}};
      \node (b)  at (0, -2)  {\lb bRy{0}};
      \node (c)  at (1.5, 0)  {\lb cRy{1}};
      \node (d)  at (1.5, -2)  {\lb dRx{0}};
      \node (e)  at (3, 0)  {\lb eWx{1}};
      \node (f)  at (4.5, 0)  {\lb fWy{1}};
    \end{scope}
		
    \path[->] (a) edge [out=225,in=135,thick] node [left=0.1cm] {dp} (b);
    \path[->] (c) edge [out=225,in=135,thick] node [left=0.1cm] {dp} (d);
    \path[->] (e) edge [out=150,in=30,dashed] node [above=0.1cm] {rf} (a);
    \path[->] (b) edge [out=-30,in=225,looseness=1.0,thick] node [below right=0.07cm] {fr} (f);
    \path[->] (f) edge [out=150,in=30,dashed] node [above=0.1cm] {rf} (c);
    \path[->] (d) edge [out=30,in=-120,thick] node [above left=0.07cm] {fr} (e);
  \end{tikzpicture}}}
  \hspace*{1.5cm}
  \subfigure[Machine model]{
  \scalebox{0.8}{
  \begin{tikzpicture}[>=stealth,thin,inner sep=0pt,text centered,shape=rectangle]
  \useasboundingbox (-0.5cm,1cm) rectangle (4cm,-3cm);
  
    \begin{scope}[minimum height=0.5cm,minimum width=0.5cm,text width=0.8cm]
      \node (da)  at (0, 0)  {$\Del(a)$};
      \node (fa)  at (0, -0.5)  {$\Flush(a)$};
      \node (db)  at (0, -2)  {$\Del(b)$};
      \node (fb)  at (0, -2.5)  {$\Flush(b)$};
      \node (dc)  at (1.5, 0)  {$\Del(c)$};
      \node (fc)  at (1.5, -0.5)  {$\Flush(c)$};
      \node (dd)  at (1.5, -2)  {$\Del(d)$};
      \node (fd)  at (1.5, -2.5)  {$\Flush(d)$};
      \node (de)  at (3, 0)  {$\Del(e)$};
      \node (fe)  at (3, -0.5)  {$\Flush(e)$};
      \node (df)  at (4.5, 0)  {$\Del(f)$};
      \node (ff)  at (4.5, -0.5)  {$\Flush(f)$};

    \end{scope}
		
    \path[->] (da) edge [out=225,in=135,thick] node [left=0.1cm] {dp} (db);
    \path[->] (fa) edge [out=225,in=135,thick] node [left=0.1cm] {se} (fb);
    \path[->] (fb) edge [out=-30,in=225,looseness=1.0,thick] node [below right=0.07cm] {se} (ff);
    \path[->] (dc) edge [out=225,in=135,thick] node [left=0.1cm] {dp} (dd);
    \path[->] (fc) edge [out=225,in=135,thick] node [left=0.1cm] {se} (fd);
    \path[->] (fa) edge [out=37,in=142,dashed,looseness=1.2] node [above=0.1cm] {de} (fe);
    \path[->] (fd) edge [out=30,in=-120,thick] node [above left=0.07cm] {se} (fe);
    \path[->] (fc) edge [out=37,in=142,dashed,looseness=1.2] node [above=0.1cm] {de} (ff);
  \end{tikzpicture}}}
\caption{\label{fig:iriw-df} Revisiting \ltest{iriw+dps} with our core machine augmented with a Power protocol}
\vspace*{-1em}
\end{figure}

In \myfig\ref{fig:iriw}, assume dependencies between the reads on $P_0$ and
$P_1$, so that $(a,b)$ on $P_0$ and $(c,d)$ on $P_1$ are safe on Power. Yet $(e,a)$ and
$(f,c)$ are delays, because Power has non-atomic writes.  Our 
machine simulates the weak behaviour exhibited on Power, following
\myfig\ref{fig:iriw-df}.  The machine enqueues or buffers all events
w.r.t.~program order.  Since $(a,b)$ and $(c,d)$ are safe on Power, our machine
augmented with a Power protocol flushes $a$ before $b$ (resp.~$c$ before $d$).
The writes corresponding to $a$ and $c$ are in the buffer, ensuring that {\tt
r1} and {\tt r3} hold the value $1$ in the end. Since $(b,f) \in \fr$
(resp.~$(d,e) \in \fr$), which is always safe, the machine flushes $b$ before
$f$ (resp.~$d$ before $e$), ensuring that $b$ and $d$ read from memory, thus
{\tt r2} and {\tt r4} hold $0$ in the end.  Finally, since $(e,a)$ and $(f,c)$
are delays, the machine flushes them in the converse order.

Formally, we establish the equivalence of the axiomatic definition of $A$
(cf.~\mysec\ref{context}) with our machine augmented with an $A$-protocol. We
write $\mns$ for \emph{machine not stuck}, \ie the machine cannot relate any
state to $\bot$ when reading a given path of labels:
\begin{theorem}\label{thm:equiv}$\valid_{A}(E,X) \wedge
\neg(\valid_{\text{SC}}(E,X)) \iff \mns(E,\wpath(E,X,\pdelays_{A}(E,X)))$
\end{theorem}

\mythm\ref{thm:equiv} shows that our machine provides a hierarchy of weak
memory models equivalent to the one of \cite{ams10}, but in an operational
style rather than an axiomatic one. 

Let us explain what intuitively matters for \mythm\ref{thm:equiv} to hold, or
in other words for us to be able to simulate a weak execution in an SC world.
A weak execution will contain at least one cycle that contradicts the
definition of SC, \eg in \myfig\ref{fig:sb} and \ref{fig:iriw}. Such a cycle
contradicts SC in that it violates program order: for example in
\myfig\ref{fig:sb}, the pair $(a,b)$ on $P_0$ is in $\po$, but there is also a
path from $b$ to $a$.

To enable SC reasoning, we need to dismantle the cycle, as illustrated in
\myfig\ref{fig:sb-df} and \myfig\ref{fig:iriw-df}. These figures recall on the
left the axiomatic cycles of \myfig\ref{fig:sb} and \ref{fig:iriw}. On the
right, they show the machine counterparts of the axiomatic cycles. We use the
following graphical conventions. In the axiomatic world (\ie on the left of our
figures), we reflect a delay pair by a dashed arrow. For example in the
\ltest{sb} test of \myfig\ref{fig:sb-df} on TSO, the write-read pairs $(a,b)$
and $(d,c)$ are delayed. In the \ltest{iriw+dps} test of
\myfig\ref{fig:iriw-df} on Power, the read-from pairs $(e,a)$ and $(f,c)$ are
delayed (as opposed to the read-read pairs $(a,b)$ on $P_0$ and $(c,d)$ on
$P_1$, which are safe thanks to the dependencies). In the machine world, the
Delayed exit rule (\ie the machine counterpart of delayed pairs) is depicted
with a dashed arrow. For safe pairs and the safe exit rule (the machine
counterpart of safe pairs), we use thick arrows, \eg the dependency $\dep$
between $a$ and $b$ on $P_0$ in \ltest{iriw+dps}.
 
First, we dismantle the axiomatic cycle by splitting an event into its delay
and flush parts, \eg the write $a$ on $P_0$ in \ltest{sb} becomes $\Del(a)$ and
$\Flush(a)$. Then, we enable SC reasoning by enforcing consistency with the
program order. For example in \myfig\ref{fig:sb-df}, the pair $(a,b)$ is in
\po~on the left, which we reflect on the right by pushing $\Del(a)$ and
$\Del(b)$ in the buffer and the queue in this order, as depicted by the
\po~arrow.

Then, the Delayed exit rule \ue~allows us to create \emph{a diversion} from the
cycle: we flush first $\Flush(b)$ then $\Flush(a)$ as depicted by the \ue~arrow
between $\Flush(b)$ and $\Flush(a)$ on the right. 

Similarly for the \ltest{iriw+dps} example recalled on the left of
\myfig\ref{fig:iriw-df}, we split all events into their $\Del$ and $\Flush$
parts, then create a diversion from the axiomatic cycle by using the Delayed 
exit rule on \eg the $(e,a)$ pair, as depicted by the \ue~arrow between
$\Flush(a)$ and $\Flush(e)$ on the right. This means that we flush the read
$\Flush(a)$ before the write $\Flush(e)$, hence the read $a$ occurs from the
queue, \ie $a$ reads the value of $e$ from the buffer.

\subsection{Reducing the Number of Delay Pairs}
\label{sec:reduced}

Crucially, the notion of creating a diversion from an SC cycle allows us to
optimise the number of pairs that we delay. \mylem\ref{lem:sc_equiv} shows that
when $(E,X)$ is SC (despite the program being run on a weak architecture $A$),
no pair needs delaying.

\paragraph{Critical delays}
 
Consider $(E,X)$ valid on $A$ but not on SC, as in
\myfig\ref{fig:sb}.  The pairs $(a,b)$ and $(c,d)$ are delays on
TSO, \ie both \emph{might} be delayed.  But it suffices to delay one of them to
reveal the weak behaviour.  Indeed, it is sufficient to buffer \eg the write
$(a)$ to $x$ on $P_0$, perform all the other events from memory (thus
$\text{\tt r1}=0$ and $\text{\tt r2}=0$ because the write $(a)$ lies in the
buffer), and finally flush the write $(a)$ to memory.

In the non-SC execution of \myfig\ref{fig:iriw}, $(e,a)$, $(f,c)$ are delays
on Power (assuming dependencies to make $(a,b)$ and $(c,d)$ safe), but it
suffices to delay \eg $(e,a)$. We buffer the write $(e)$, enqueue the read
$(a)$, and perform other events from memory. This corresponds to a caching
scenario where $P_0$ and $P_3$ communicate privately \emph{via} $x$ (since they
communicate from the buffer to the queue), all the other communications
occurring from memory, in an SC fashion.
 
Note that in both cases, the reasoning would have been similar if we had chosen
another delay pair along the cycle exhibited by the execution, or if
we had delayed more events than just the ones we chose. As said before, what
matters is to delay enough pairs to form a diversion from the cycle, to
enable an SC reasoning. 

We said before that when an execution is SC, no pair needs delaying.
\cite[Thm.1]{am11} characterises the non-SC executions by the presence of
certain cycles, called \emph{critical cycles}, which satisfy the two following
conditions. {\bf (i)}~Per processor, the cycle involves at most two memory
accesses $a$ and $b$ on this processor and $\loc(a) \neq \loc(b)$.
{\bf (ii)}~For a given memory location $x$, the cycle involves at most three accesses
relative to x, and these accesses are from distinct processors. The executions
of the tests \ltest{sb} and \ltest{iriw} in \myfig\ref{fig:sb} and
\ref{fig:iriw} give typical examples of critical cycles. 

Thus, if there is a weak memory specific bug in an execution, it is along a
critical cycle, or on the remainder of a path after a critical cycle.  
Formally, we show that we only need to delay (\ie apply the Delayed
exit rule \ue) \emph{one} delay pair per critical cycle to simulate a weak
execution with our machine (writing $\cdelays_{A}(E,X)$ for \emph{any}
selection of pairs in $\pdelays_{A}(E,X)$ with at least one pair per
critical cycle of $E$):
\begin{theorem}\label{thm:optim}
($\valid_{A}(E,X) \wedge \neg(\valid_{\text{SC}}(E,X))) \iff \mns(E,\wpath(E,X,\cdelays_{A}(E,X)))$
\end{theorem}

\mythm\ref{thm:optim} has several consequences. From the semantic
perspective, it defines a family of paths equivalent to $(E,X)$, \ie the
paths built inductively as above from any selection of critical pairs.  Thus
one can see the partial orders given by $(E,X)$ in the axiomatic model as
the canonical representation of all the equivalent operational paths.

From the verification perspective, the critical pairs highlight the
instructions that should be delayed when verifying a program running on weak
memory.  Crucially, we show that only one delay pair per critical cycle
actually needs delaying.  This enables efficient verification, as shown by our
experiments (cf.~in \mysec\ref{experiments} the time taken by SatAbs to verify
the two versions of the PostgreSQL excerpt: $21.34$ vs.~$1.29$ seconds).

\paragraph{Transformation}

Consider a non-SC execution and a selection of delays, \eg the execution of
\ltest{sb} in \myfig\ref{fig:sb-df} and the pair $(a,b)$. As said before, we
only need to create a diversion from the cycle, here by using the Delay exit
rule \ue{} on $(a,b)$, \ie flushing $\Flush(b)$ then $\Flush(a)$ as depicted on
the right of \myfig\ref{fig:sb-df}. Consider now that the other pairs, in our
example $(c,d)$, are not delayed, in the sense that we use the Safe exit rule
to handle them; this amounts to having an \se{}~arrow between $\Flush(c)$ and
$\Flush(d)$. This scenario corresponds to a situation where the write $c$
writes directly to memory (\ie in our machine writes to the buffer but is
flushed immediately after), and the read $d$ reads from memory as well.

Thus in practice, we will tag an event with $\Mem$ to mean that we perform it
w.r.t.~memory. Otherwise, given a selection $\mathbb{D}$ of delays, we delay an event $e$
(\ie tag it $\Del$) when:
\begin{description}
\item[Source of program order \dpo] there is an event $e'$ after $e$ in program order
(\ie $(e,e') \in \po$), forming a delay pair with $e$ (\ie $(e,e') \in
\mathbb{D})$; or
\item[Source of read-from \drfs] there is a read $e'$ from another thread
reading from $e$ (\ie $(e,e') \in \rfe$), forming a delay pair with $e$ (\ie
$(e,e') \in \mathbb{D})$; or
\item[Target of read-from \drft] there is a write $e'$ from another thread from
which $e$ reads (\ie $(e',e) \in \rfe$), forming a delay pair with $e$ (\ie
$(e',e) \in \mathbb{D}$).
\end{description}

Thus, we simulate a non-SC execution on an architecture $A$ as follows: 1.~We find
the critical cycles.  2.~We select at least one delay pair per critical cycle.
3.~We tag the events in these pairs with $\Del$ w.r.t.~\dpo, \drfs~and \drft,
and all the others with $\Mem$.  4.~We perform the events tagged $\Del$ from
the buffer or the queue, and the events tagged $\Mem$ from memory.  5.~We flush
the delay pairs following the Delayed exit rule \ue, and the other pairs
following the Safe exit rule \se.
\section{Implementation\label{implementation}}

We implemented the transformation technique described above.  Our tool reads a
concurrent C program, and generates a new concurrent C program augmented with
buffers and queues, which is then passed to an SC verification tool.  We added
two memory fences ({\tt fence} and {\tt lwfence}) as new C keywords to support
x86's {\tt mfence} and Power's {\tt sync} and {\tt lwsync} with the semantics
presented in~\mysec\ref{context}. 

We first translate the C source code into a \emph{goto-program} (a control
flow graph), then feed it to \prog{goto-instrument} (a tool
automating transformations of goto-programs).  We have extended
\prog{goto-instrument} with the transformation described in
\mysec\ref{instrumentation}: given a memory model (x86/TSO, PSO, RMO and Power
are available for now), \prog{goto-instrument} adds the instructions necessary
to transform the delay pairs. 

\begin{figure}[!t]
\begin{tabular}{m{.6\linewidth}@{}m{.4\linewidth}}
\begin{lstlisting}[morekeywords={end, assert, thd1, thd2, thd3, thd4, main},escapechar=@]
thd1:             thd2:           thd3:   thd4: 
@\underline{r1:=x}@             @\underline{r3:=y}@           @\,\underline{x:=1}@    @\underline{y:=1}@   
tmp1:=x xor x     tmp2:=y xor y                
r2:=y+tmp1        r4:=x+tmp2                    

main:
thd1() || thd2() || thd3() || thd4()
assert(!(r1=1 & r2=0 & r3=1 & r4=0))
\end{lstlisting}
&
  \scalebox{0.7}{
  \begin{tikzpicture}[>=stealth,thin,inner sep=0pt,text centered,shape=rectangle]
  \useasboundingbox (-1.5cm,1cm) rectangle (3.5cm,-3.2cm);

    \begin{scope}[minimum height=0.5cm,minimum width=0.5cm,text width=1.0cm]
      \node (a)  at (0, 0)  {\lb a\underline{Rx}{}};
      \node (b)  at (0, -2)  {\lb bRy{}};
      \node (c)  at (1.5, 0)  {\lb c\underline{Ry}{}};
      \node (d)  at (1.5, -2)  {\lb dRx{}};
      \node (e)  at (3.0, 0)  {\lb e\underline{Wx}{}};
      \node (f)  at (4.5, 0)  {\lb f\underline{Wy}{}};
    \end{scope}

    \path[->] (a) edge [out=225,in=135] node [left=0.1cm] {dp} (b);
    \path[->] (c) edge [out=225,in=135] node [left=0.1cm] {dp} (d);
    \path[<->] (e) edge [out=150,in=30] node [above=0.1cm] {} (a);
    \path[<->] (b) edge [out=-30,in=225,looseness=1.0] node [below right=0.07cm] {} (f);
    \path[<->] (f) edge [out=150,in=30] node [above=0.1cm] {} (c);
    \path[<->] (d) edge [out=30,in=-120] node [above left=0.07cm] {} (e);
  \end{tikzpicture}}
\end{tabular}
\vspace*{-3mm}
\hrule
\begin{lstlisting}[morekeywords={end, assert, thd1, thd2, thd3, thd4, main, atomic},escapechar=@]
thd1:                          thd3:                      main:
// was r1:=x                   // was x:=1                @\text{[...]}@
if(*)                          if(*)                      // was assert
  r1:=buff_x.take(thd1)          buff_x.push(1,thd3)      if(!(delay_r1=0) & *)
else                           else                         r1:=deref(delay_r1)
  delay_r1:=ref(x) end           x:=1 end                   delay_r1:=0 end
[...]                                                     // same for delay_r2
                                                          assert(!(r1=1 & r2=0
// same for thd2               // same for thd4                  @\text{\& r3=1 \& r4=0))}@
\end{lstlisting}
\vspace*{-3mm}
\caption{\label{fig:iriw+dps} Study of transformation of \ltest{iriw+dps}}
\vspace*{-4mm}
\end{figure}

Let us explain how we transform a program using the example \ltest{iriw+dps}
(a variation of the \ltest{iriw} test of \myfig\ref{fig:iriw}, augmented with
dependencies between the read pairs to make these pairs safe). We give this
program (in C) at the top of \myfig\ref{fig:iriw+dps}.

To transform \ltest{iriw+dps}, \prog{goto-instrument} first produces the
abstract event graph on the right-hand side of \myfig\ref{fig:iriw+dps} from
the control-flow graph of the program, which over-approximates the event
structures by ignoring the values of the variables, and each abstract event may
correspond to an unbounded number of concrete events.
The tool then computes possible critical cycles on abstract events. By taking
into account any possible concretisation of these events, these cycles are an
over-approximation of the actual cycles present in concrete executions of
this program, but spurious cycles do not impair
the soundness of our method. On our example, due to the direction of $\dep$
arrows, the only possible cycle corresponds to the one in \myfig\ref{fig:iriw}:
$a, b, f, c, d, e, a$.

Next, \prog{goto-instrument} computes the delay pairs for the selected memory model (here, for
Power, $(e,a)$, and $(f,c)$). Following \mysec\ref{sec:reduced},
we can transform \emph{all} the delay pairs (following
\mythm\ref{thm:equiv}), or only one pair per cycle (following
\mythm\ref{thm:optim}).  As we discuss in \mysec\ref{experiments}, this choice
can have a drastic impact on the time required by the program analyser (\eg
$21.34$ vs.~$1.29$ seconds for SatAbs on the PostgreSQL excerpt).

The tool then transforms these pairs (underlined instructions and events in
\myfig\ref{fig:iriw+dps}) using a buffer of size~$2$ for each variable. To
ensure soundness despite this limitation, it adds assertions to check whether
this buffer bound is exceeded.

The lower part of \myfig\ref{fig:iriw+dps} gives an excerpt of the resulting
transformed program (only the first read of {\tt thd1}, and the write of {\tt thd3}).
%
%
We transform a write into {\tt x} (\eg~{\tt x:=1} in {\tt thd3}) with a
non-deterministic choice ({\tt if(*)}): either the write directly hits the memory, as
it would without transformation, or the write is stored into the buffer ({\tt buff\_x.push(1,thd3)}).
%
We transform a load from {\tt x} into register {\tt r1} (\eg~{\tt r1:=x} in {\tt
thd1}) with another non-deterministic choice: either a load from the memory or
from the write buffer of {\tt x} ({\tt r1:=buff\_x.take(thd1)}), or a postponed
read ({\tt delay\_r1:=ref(x)}). 

Postponing a read models the read entering the queue. In this case, we do not
update {\tt r1}, but the pointer (initially null) {\tt delay\_r1} is set to {\tt x}
({\tt delay\_r1:=ref(x)} where {\tt ref(x)} returns the address of {\tt x}).  This
pointer 1) states that the value of {\tt r1} might be affected by a delayed
read in the queue (as it is not null anymore), and 2) keeps track of the
variable {\tt x}, which can be read by a read in the queue. When a subsequent
instruction reads the value of {\tt r1} (\eg~{\tt assert(!(r1 \&
...))} in function {\tt main}) and {\tt delay\_r1} is null, then we read the current value
of {\tt r1} because there is no delayed read in the
queue affecting {\tt r1}.  If {\tt delay\_r1} points to a variable {\tt x}, then
there are two cases (again a non-deterministic choice).  Either the flush of
the read of {\tt x} in the queue happens after the {\tt assert}, in which case
{\tt assert} reads the current value of {\tt r1}.  Or the flush happens before, 
in which case we update {\tt r1} with the current value in memory of the
variable pointed to by
{\tt delay\_r1}, namely {\tt x} (\emph{via} {\tt r1:=deref(delay\_r1)}, where {\tt
deref} dereferences the pointer in argument).  
%



The transformed goto-program can be given directly to CImpact or SatAbs.
Alternatively, we can convert it back into C code, and hand it to any program
analyser that can read C source (\eg CheckFence, ESBMC, Poirot, Threader). 
We also wrote a converter to Blender's input format, and a C\# generator for
MMChecker.


\section{Experimental Results \label{experiments}}

We exercised our method and measured its cost using~$8$ tools.  We
considered~$5$ ANSI-C model checkers: SatAbs, a verifier based on predicate
abstraction, using Boom as the model checker for the Boolean
program; ESBMC, a bounded model checker; CImpact, a variant of the Impact
algorithm extended to SC concurrency; Threader, a thread-modular
verifier; and Poirot, which implements a context-bounded translation to
sequential programs.  These tools cover a broad spectrum of
symbolic algorithms for verifying SC programs.  We also experimented with
Blender, CheckFence, and MM\-Checker.  We ran our experiments on
Linux 2.6.32 64-bit machines with 3.07~GHz (only Poirot was run
on a Windows system).

\paragraph{Validation}

First, we systematically validate our setup using $555$ litmus tests exposing
weak memory artefacts (\eg instruction reordering, store buffering, write
atomicity relaxation) in isolation. The \prog{diy} tool automatically generates
x86, Power and ARM assembly programs implementing an idiom that cannot be
reached on SC, but can be reached on a given model.  For example, the
\ltest{sb} test of \myfig\ref{fig:sb} exhibits store buffering, thus can be
reached on any weak model, from TSO to Power. The \ltest{iriw} test of
\myfig\ref{fig:iriw} can only be reached on RMO (by reordering the reads) or on
Power (for the same reason, or because the writes are non-atomic). Finally,
\ltest{iriw+dps} (\ie \ltest{iriw} with dependencies between the reads to
prevent their reordering) can only be reached on Power.

\begin{figure}[!t]
\centering
\subfigure[Average of SatAbs on selected litmus tests]{
\includegraphics[width=0.46\textwidth]{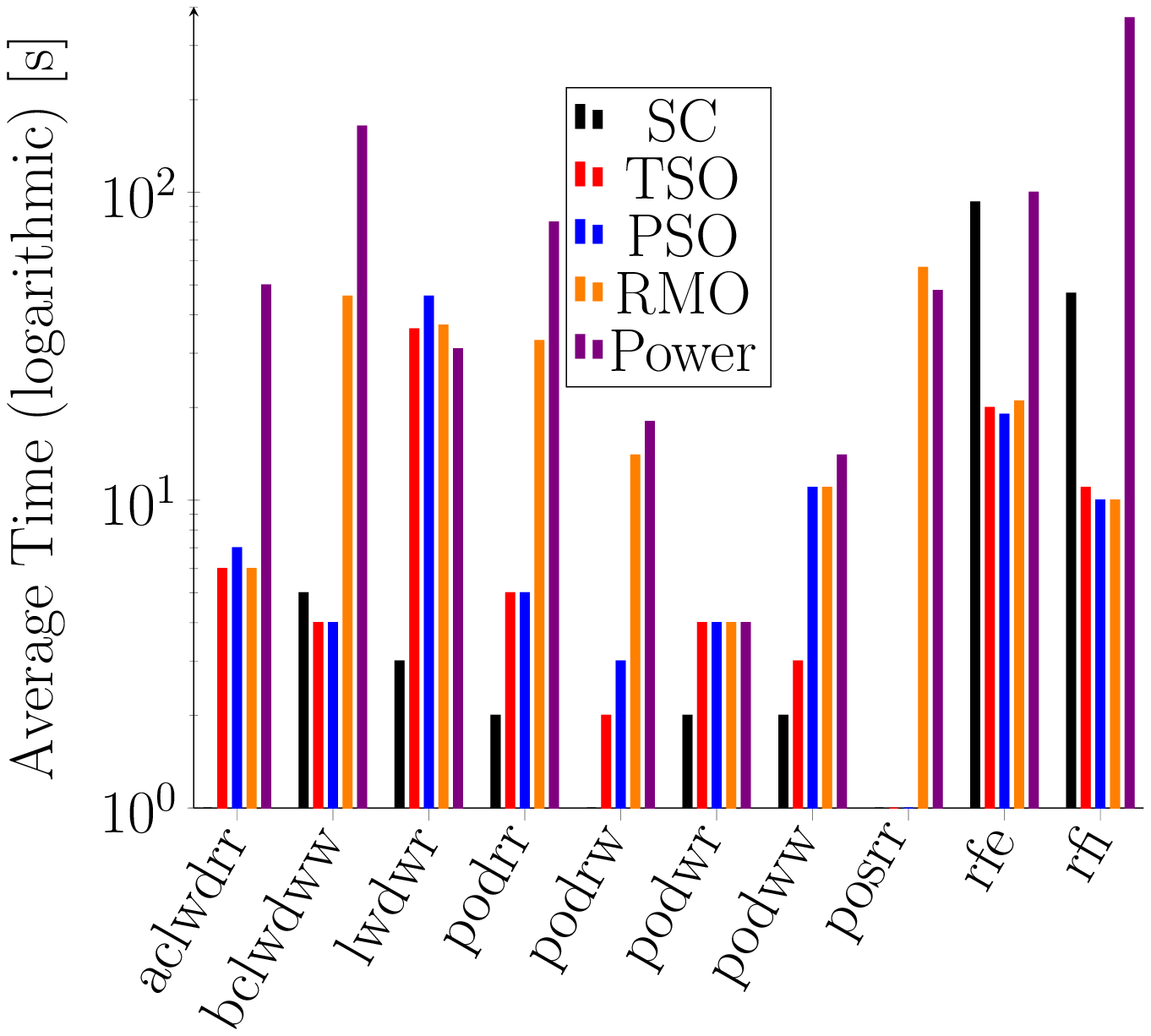}
\label{fig:satabs-average}
}
  \subfigure[SatAbs on SC, Power full and optimised]{
\includegraphics[width=0.5\textwidth]{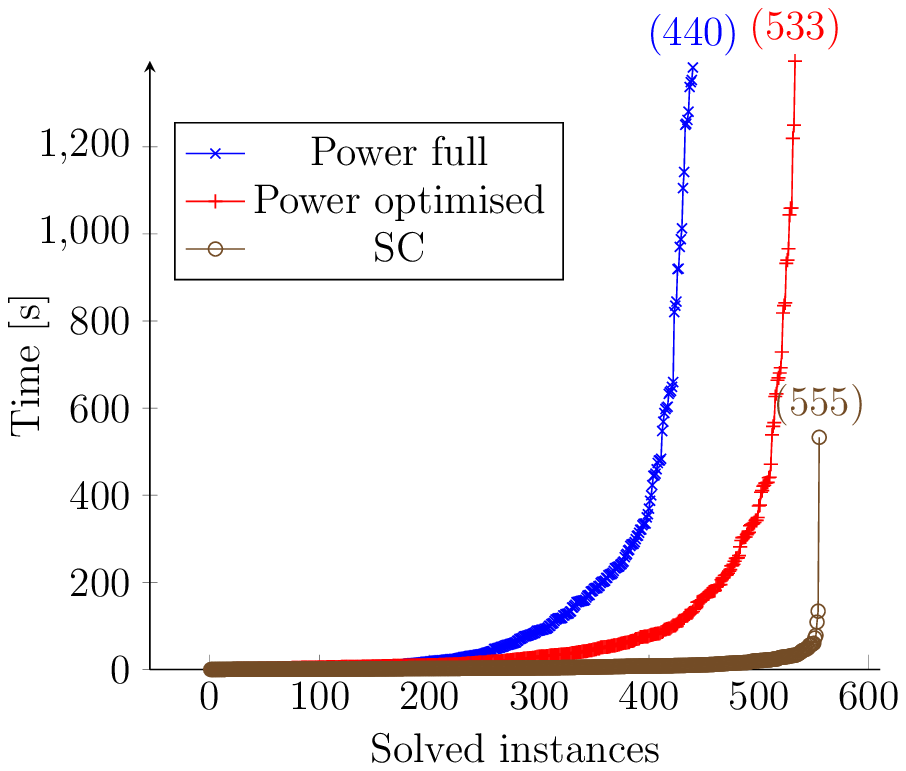}
\label{fig:satabs-power}
}
\vspace*{-2mm}
\caption{Selected experimental results}
\label{fig:experiments}
\vspace*{-3mm}
\end{figure}

\begin{wrapfigure}{r}{6.8cm}
  \vspace*{-2em}
  \hspace*{-2mm}
  \includegraphics[width=7cm]{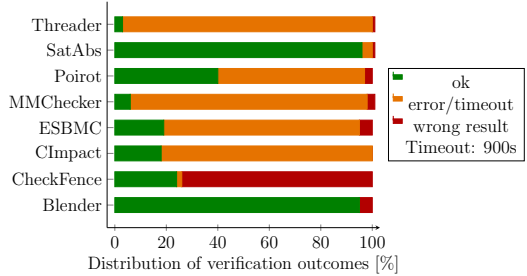}
\vspace*{-2em}
\caption{Comparison of tools on all tests and models}
\vspace*{-2em}
\label{fig:all-tools}
\end{wrapfigure}

Each litmus test comes with an assertion expressing the SC violation exercised
by the test, \eg the outcomes of \myfig\ref{fig:sb} and \ref{fig:iriw}. Thus,
verifying a litmus test amounts to checking whether the model under scrutiny
can reach the specified outcome. We then convert these tests automatically into
C code, leading to programs of $48$ lines on average, involving $2$ to $4$ threads.

These examples allow us to check that we soundly implement the theory of
\mysec\ref{instrumentation}: we verify each test w.r.t.~SC, \ie no
transformation, then w.r.t.~TSO, PSO, RMO, and Power.  Despite the tests being
small, they provide challenging concurrent idioms to verify.
\myfig\ref{fig:all-tools} compares the tools on all tests and models.  Most
tools, with the exception of Blender and SatAbs, timeout or give wrong results on
a vast majority of tests.

\myfig\ref{fig:satabs-average} gives the average time that SatAbs needs to
verify several litmus families (\eg rfe tests exercise store atomicity, podwr
tests exercise the write-read reordering), for all tools and models, from SC to
Power. We also compare the full (\mysec\ref{sec:full}) and optimised
transformation (\mysec\ref{sec:reduced}) for all models, tools and tests.
Thus each run consists of $9 \times 555$ distinct instances.
\myfig\ref{fig:satabs-power} shows that the optimised approach allows SatAbs to
verify $533$ of the $555$ tests, and $440$ with the full approach, in more than
twice the time needed for SC. We give the results for all experiments online.  

We also verified several TSO examples (details are online). Note that these
examples in fact only exhibit idioms already covered by our litmus tests (\eg
Dekker corresponds to the \ltest{sb} test of \myfig\ref{fig:sb}). We now study
a real-life example, an excerpt of the relational database software PostgreSQL.

%
%


\paragraph{Worker Synchronization in PostgreSQL}
Mid 2011, PostgreSQL developers observed that a regression test occasionally
failed on a multi-core PowerPC system.%
\footnote{http://archives.postgresql.org/pgsql-hackers/2011-08/msg00330.php}
The test implements a protocol passing a token in a ring of processes.  Further
analysis drew the attention to an interprocess signalling mechanism.  It turned
out that the code had already been subject to an inconclusive
discussion in late 2010.%
\footnote{http://archives.postgresql.org/pgsql-hackers/2010-11/msg01575.php}

\begin{wrapfigure}{l}{5.2cm}
\vspace*{-2.5em}
\begin{lstlisting}[caption={C source code of token passing}, label={prog:pgsql}]
#define WORKERS 2
volatile _Bool latch[WORKERS];
volatile _Bool flag[WORKERS];
void worker(int i) 
{ while(!latch[i]);
  for(;;) 
  { assert(!latch[i] || flag[i]); (*@ \label{pgsql:starve} @*)
    latch[i] = 0; (*@ \label{pqsql:setlatch} @*)
    if(flag[i]) (*@\label{pgsql:dowork} @*)
    { flag[i] = 0; 
      flag[(i+1)%WORKERS] = 1; (*@\label{pgsql:nextflag}@*)
      latch[(i+1)%WORKERS] = 1; (*@\label{pgsql:nextlatch}@*) } 
    while(!latch[i]); (*@ \label{pgsql:waitlatch} @*) } }
\end{lstlisting}
\vspace*{-2.5em}
\end{wrapfigure}

The code in Listing~\ref{prog:pgsql} is an inlined version of the
problematic code, with an additional assertion in line~\ref{pgsql:starve}. 
Each element of the array ``\lstinline{latch}'' is a Boolean variable stored
in shared memory to facilitate interprocess communication.  Each working
process waits to have its latch set and then expects to have work to do
(from line~\ref{pgsql:dowork} onwards).  Here, the work consists of passing
around a token \emph{via} the array ``\lstinline{flag}''.  Once the process
is done with its work, it passes the token on (line~\ref{pgsql:nextflag}),
and sets the latch of the process the token was passed to
(line~\ref{pgsql:nextlatch}).

Starvation seemingly cannot occur: when a process is woken up, it has work to
do (has the token). Yet, the PostgreSQL developers observed that the wait in
line~\ref{pgsql:waitlatch} (which in the original code is bounded in time)
would time out, thus signalling starvation of the ring of processes.  Manual
inspection identified the memory model of the platform as possible culprit: it
was assumed that the processor would at times delay the write in
line~\ref{pgsql:nextflag} until after the latch had been set.

\medskip

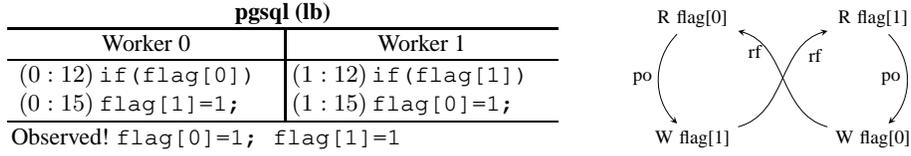
\begin{figure}[!t]
\begin{tabular}{m{.6\linewidth}m{.6\linewidth}}
\centerline{
\begin{tabularx}{\linewidth}{Y|Y}
\multicolumn{4}{c}{\ltest{pgsql (lb)}} \\ \hline
\multicolumn{2}{c|}{\haut Worker {0}} &
\multicolumn{2}{c}{\haut Worker {1}} \\ \hline
\haut\instab{if(flag[0])}{(0:12)} & \instab{if(flag[1])}{(1:12)} \\
\bas\instab{flag[1]=1;}{(0:15)} & \instab{flag[0]=1;}{(1:15)} \\ \hline
\multicolumn{4}{l}{\haut{}Observed! \as{flag[0]=1; \as{flag[1]=1}}}
\end{tabularx}
}
&
\centerline{
  \scalebox{0.8}{
  \begin{tikzpicture}[>=stealth,thin,inner sep=0pt,text centered,shape=rectangle]
  \useasboundingbox (1cm,-1.8cm) rectangle (4cm,0cm);
    \begin{scope}[minimum height=0.5cm,minimum width=0.5cm,text width=1.5cm]
      \node (a)  at (0, 0)  {{R flag[0]}};
      \node (b)  at (0, -2)  {W flag[1]};
      \node (c)  at (3, 0)  {R flag[1]};
      \node (d)  at (3, -2)  {W flag[0]};
    \end{scope}
    \path[->] (a) edge [out=225,in=135] node [left=0.1cm] {po} (b);
    \path[->] (b) edge [out=15,in=195]  node [pos=0.8,below right=0.07cm] {rf} (c);
    \path[->] (c) edge [out=-45,in=45]  node [left=0.1cm] {po} (d);
    \path[->] (d) edge [out=165,in=-15] node [pos=0.8,below left==0.07cm] {rf} (a);
  \end{tikzpicture}}
}
\end{tabular}
\vspace*{-8mm}
\caption{\label{fig:pgsql-lb} An \ltest{lb} idiom detected in {\tt pgsql.c}}
\vspace*{-4mm}
\end{figure}

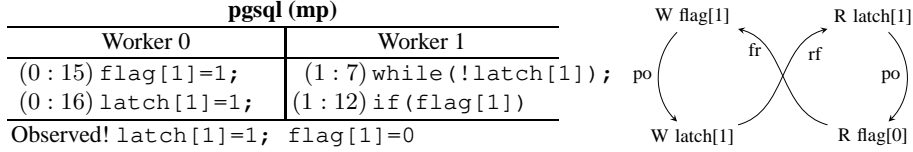
\begin{figure}[!t]
\begin{tabular}{m{.6\linewidth}m{.6\linewidth}}
\centerline{
\begin{tabularx}{\linewidth}{Y|Y}
\multicolumn{4}{c}{\ltest{pgsql (mp)}} \\ \hline
\multicolumn{2}{c|}{\haut Worker {0}} &
\multicolumn{2}{c}{\haut Worker {1}} \\ \hline
\haut\instab{flag[1]=1;}{(0:15)} & \instab{while(!latch[1]);}{(1:7)} \\
\bas\instab{latch[1]=1;}{(0:16)} & \instab{if(flag[1])}{(1:12)} \\ \hline
\multicolumn{4}{l}{\haut{}Observed! \as{latch[1]=1; \as{flag[1]=0}}}
\end{tabularx}
}
&
\centerline{
  \scalebox{0.8}{
  \begin{tikzpicture}[>=stealth,thin,inner sep=0pt,text centered,shape=rectangle]
  \useasboundingbox (1cm,-1.8cm) rectangle (4cm,0cm);
    \begin{scope}[minimum height=0.5cm,minimum width=0.5cm,text width=1.5cm]
      \node (a)  at (0, 0)  {{W flag[1]}};
      \node (b)  at (0, -2)  {W latch[1]};
      \node (c)  at (3, 0)  {R latch[1]};
      \node (d)  at (3, -2)  {R flag[0]};
    \end{scope}
    \path[->] (a) edge [out=225,in=135] node [left=0.1cm] {po} (b);
    \path[->] (b) edge [out=15,in=195]  node [pos=0.8,below right=0.07cm] {rf} (c);
    \path[->] (c) edge [out=-45,in=45]  node [left=0.1cm] {po} (d);
    \path[->] (d) edge [out=165,in=-15] node [pos=0.8,below left==0.07cm] {fr} (a);
  \end{tikzpicture}}
}
\end{tabular}
\vspace*{-8mm}
\caption{\label{fig:pgsql-mp} An \ltest{mp} idiom detected in {\tt pgsql.c}}
\vspace*{-4mm}
\end{figure}

We transform the code of Listing~\ref{prog:pgsql} for two workers under Power.
The \prog{goto-instrument} graph shows two idioms: \ltest{lb} (load buffering)
and \ltest{mp} (message passing), in \myfig\ref{fig:pgsql-lb} and
\ref{fig:pgsql-mp}, specifying the corresponding lines of
Listing~\ref{prog:pgsql}.

The \ltest{lb} idiom contains the two \emph{if} statements controlling the
access to both critical sections. Since the \ltest{lb} idiom is yet
unimplemented by Power machines, (despite being allowed by the
architecture~\cite{ssa11}), we believe that this is not the bug observed by the
PostgreSQL developers. Yet, it might lead to actual bugs on future machines.

By contrast, the \ltest{mp} case is commonly observed on Power machines (\eg
1.7G/167G on Power $7$ \cite{ssa11}). The \ltest{mp} case arises in the
PostgreSQL code by the combination of some writes in the critical section of
the first worker, and the access to the critical section of the second worker,
which lines we give in \myfig\ref{fig:pgsql-mp}.

We first check the fully transformed code with SatAbs. After 21.34 seconds,
SatAbs provides a counterexample (given online), where we first execute the
first worker up to the line 17.  All accesses are w.r.t.~memory, except at
lines~14 and 15, where the values 0 and 1 are stored into the buffers of
flag[0] and flag[1].  Then the second worker starts, reading the updated
value $1$ of latch[1].  It exits the blocking while (line 7) and reaches the
assertion.  Here, latch[1] still holds $1$, and flag[1] still holds $0$, as
Worker $0$ has not flushed yet the write waiting in its buffer.  Thus, the
condition of the \emph{if} is not true, the critical section is skipped, and
the program arrives line 19, without having authorised the next worker to
enter in critical section, and loops forever.

As \ltest{mp} can arise on Power \eg because of non-atomic writes, we know by
\mythm\ref{thm:optim} that we only need to transform one \rfe~pair of the
cycle, and relaunch the verification. SatAbs spends 1.29 second to check it
(and finds a counterexample, as previously).

PostgreSQL developers had discussed ways of fixing this, but only committed
comments to the code base as it remained unclear whether the intended fixes
were appropriate. We proposed a provably correct patch solving both \ltest{lb}
and \ltest{mp}.  After discussion with the
developers\footnote{http://archives.postgresql.org/pgsql-hackers/2012-03/msg01506.php},
we improved it to meet the developers' desire to maintain the current API. The
final patch places two {\tt lwsync} barriers: after line~\ref{pqsql:setlatch}
and before line~\ref{pgsql:nextlatch}.

\section{Conclusion}

We presented a provably sound method to verify concurrent software w.r.t.~weak
memory. Our contribution allows to lift SC methods and tools to a wide range of
weak memory models (from x86 to Power), by the mean of program transformation.

Our approach crucially relies on the definition of a generic operational model
equivalent to the axiomatic one of~\cite{ams10}. We do not favor any style of
model in particular, but we highlight the importance of having several
equivalent mathematic styles to describe a field as intricate as weak memory.
In addition, operational models are often the style of choice in the
verification community; we contribute here to the vocabular to tackle the
verification problem w.r.t.~weak memory. 

Our extensive experiments and in particular the PostgreSQL bug demonstrate the
practicability of our approach from several different perspectives. First, we
confirmed an existing bug (\ltest{mp}), and validated the fix proposed by the
developers, including evaluation of different synchronisation options. Second,
we found an additional idiom (\ltest{lb}), which will be a bug on future Power
machines; our fix repairs it already. Third, our work raised notable interest
in both the open-source world (see our discussion with the PostgreSQL
developers) and in industry (our nascent collaboration with IBM).  The
verification problem under weak memory is far from being solved (in particular
for scalability reasons, as shown by our experiments) but we made a convincing
first step.

\bibliographystyle{plain}
\renewcommand{\baselinestretch}{0.86}
\bibliography{bibliography,jade}


\end{document}